# THE PREDICTION OF MASS OF $Z'$-BOSON FROM $B_q^0 - \overline{B_q^0}$ MIXING

## S. Sahoo[1], C. K. Das[2] and L. Maharana[3]

[1]Department of Physics, National Institute of Technology,
Durgapur-713209, West Bengal, India
E-mail: sukadevsahoo@yahoo.com

[2]Department of Physics, Trident Academy of Technology,
Bhubaneswar – 751 024, Orissa, India.

[3]Department of Physics, Krupajal Engineering College,
Bhubaneswar, Orissa, India.

**ABSTRACT**

$B_q^0 - \overline{B_q^0}$ mixing offers a profound probe into the effects of new physics beyond the Standard Model. In this paper, $B_s^0 - \overline{B_s^0}$ and $B_d^0 - \overline{B_d^0}$ mass differences are considered taking the effect of both Z- and $Z'$-mediated flavour-changing neutral currents in the $B_q^0 - \overline{B_q^0}$ mixing (q = d, s). Our estimated mass of $Z'$ boson is accessible at the experiments LHC and B-factories in near future.

## 1. Introduction

One of the most promising ways to detect the effects of new physics (NP) on *B* decays is to look for deviations of flavour-changing neutral current (FCNC) processes from their Standard Model (SM) predictions.[1] FCNC process occurs at loop-level in the SM. Its rate is suppressed by small electroweak gauge coupling, CKM matrix elements and loop factors.[2] On one hand these processes are very sensitive probe of NP beyond the SM because some of these suppression factors can be enhanced in NP models. On the other hand FCNC processes of *K*, $B_d$ and $B_s$ mesons[3] are still large enough to be studied experimentally as well as theoretically. $B_q^0 - \overline{B_q^0}$ mixing (q = d, s), meson-antimeson mixing,[4–6] plays an outstanding role in this direction. First, meson-antimeson oscillations occur at time scales which are sufficiently close to the meson lifetimes to permit their experimental investigation. Second, the SM contribution to meson-antimeson mixing is loop-suppressed and comes with two or more small elements of the CKM matrix.[7,8] Third, the decays of oscillating mesons give access to many mixing-induced CP asymmetries through the time-dependent study of decays into CP-eigenstates, which in some cases one can relate to the parameters of the underlying theory with negligible hadronic uncertainties.

In $B_q^0 - \overline{B_q^0}$ mixing, an initially present $B_q^0$ state evolves into a time-dependent linear combination of $B_q^0$ and $\overline{B_q^0}$ flavour states. The oscillation frequency of this phenomenon is characterized by the mass difference of the 'heavy' and 'light' mass eigenstates:

$$\Delta M_{B_q} \equiv M_H(B_q) - M_L(B_q) = 2|M_{12}(B_q)|. \tag{1}$$

The determination of $B_q^0 - \overline{B_q^0}$ mass difference $\Delta M_{B_q}$ has been a major objective of particle physics. The phenomenon of $B_d^0$ oscillations is well established,[9] with a precisely measured mass difference $\Delta M_{B_d}$. In the SM, this parameter is proportional to the combination $(V_{td}^* V_{tb})^2$ of CKM matrix elements. Since the matrix element $V_{ts}$ is larger than $V_{td}$, the expected mass difference $\Delta M_{B_s}$ is higher. Hence, the mass differences $\Delta M_{B_d}$ and $\Delta M_{B_s}$ can be used to determine CKM matrix elements $V_{td}$ and $V_{ts}$ respectively, which relates the quark mass eigenstates to the flavour eigenstates. In the SM[10,11] with 3–$\sigma$ range, $B_s^0 - \overline{B_s^0}$ and $B_d^0 - \overline{B_d^0}$ mass differences are found to be:

$$(\Delta M_{B_d})_{SM} = 0.394^{+0.361}_{-0.162} \text{ ps}^{-1}, \tag{2}$$

$$(\Delta M_{B_s})_{SM} = 21.7^{+13.1}_{-9.1} \text{ ps}^{-1}, \tag{3}$$

From the recent experiments, $B_s^0 - \overline{B_s^0}$ and $B_d^0 - \overline{B_d^0}$ mass differences are found to be:

$$\Delta M_{B_d} = 0.507 \pm 0.005 \text{ ps}^{-1} \quad [\text{ref. 4}], \tag{4}$$

$$\Delta M_{B_s} = 17.77 \pm 0.10 \, (stat) \pm 0.07 \, (syst) \text{ ps}^{-1} \; (CDF) \quad [\text{ref. 12}], \tag{5}$$

and $\quad 17 < \Delta M_{B_s} < 21 \text{ ps}^{-1} \; (90\% \text{ CL}) \; (D\emptyset) \quad\quad [\text{ref. 13}]. \tag{6}$

Although these experimental values are a little bit different from their SM values, for large hadronic uncertainties we can not strongly argue that it is a NP signal. However, these measurements may give constraints on the NP models, which predict $b \to s(d)$ FCNC transitions. This is why the $B_q^0 - \overline{B_q^0}$ mixing is one of the most important and interesting portals for detection of NP models.[14]

The $Z'$ is a hypothetical massive, electrically-neutral spin 1 gauge boson.[4] These bosons are predicted by a wide variety of extensions of the SM.[15–21] Theoretically it is predicted that they exist in Grand Unified Theories (GUTs), left-right symmetric models, Little Higgs models, superstring theories and theories with large extra dimensions. But experimentally $Z'$ boson is not conclusively discovered so far. Hence, the exact mass of $Z'$ boson is not known. The current experimental searches of the $Z'$ boson from Drell-Yan cross sections at Tevatron have put lower limits on the mass range 0.6 – 1.0 TeV at 95 % C. L. depending on the specific models.[22] However, the lower mass limit can be as low as[23] 130 GeV if the coupling is weak. For an experimentalist a $Z'$ is a resonance 'bump' more massive than the Z of the SM which can be observed in Drell-Yan production followed by its decay into lepton-antilepton pairs.[24] For a phenomenologist a $Z'$ boson is a new massive electrically neutral, colourless boson (equal to its own antiparticle) which couples to SM matter. For a theorist it is useful to classify the $Z'$ according to its spin, even though actually measuring its spin will require high statistics.

There are many models beyond the SM predict more than one extra neutral gauge bosons and many new fermions. These new (exotic) fermions can mix with the SM fermions. Such mixing induces FCNCs.[25,26] Mixing between ordinary (doublet) and exotic singlet left-handed quarks induces FCNC, mediated by the SM Z boson. In these models[27–29], one introduces an additional vector-singlet charge –1/3 quark h, and allows it to mix with the ordinary down-type quarks d, s and b. Since the weak isospin of the exotic quark is different from that of the ordinary quarks, FCNCs involving Z are induced. The Z-mediated FCNC couplings $U_{ds}^Z$, $U_{db}^Z$ and $U_{sb}^Z$ which are in general complex, are constrained by a variety of processes. $U_{ds}^Z$ is bounded by the measurements of $\Delta M_K$ ($K^0$-$\overline{K}^0$ mixing), $|\epsilon|$ (the CP-violating parameters in the kaon system) and $K_L \to \mu^+\mu^-$,[27–29] while the constraints on $U_{db}^Z$ and $U_{sb}^Z$ come principally from the experimental limit on B $\left(B \to \ell^+\ell^- X\right)$.[30–33] The constraints on $U_{db}^Z$ and $U_{sb}^Z$ allow significant contributions to $B_q^0 - \overline{B_q^0}$ mixing (q = d, s). Models of NP, which contain exotic fermions also predict the existence of additional neutral $Z'$ gauge bosons. The mixing among particles which have different $Z'$ quantum numbers will induce FCNCs due to $Z'$ exchange.[34,35] With FCNCs, the $Z'$ boson contributes at tree level, and its contribution will interfere with the SM contributions.

In this paper, $B_s^0 - \overline{B_s^0}$ and $B_d^0 - \overline{B_d^0}$ mass differences are considered taking the effect of both Z- and $Z'$-mediated FCNCs in the $B_q^0 - \overline{B_q^0}$ mixing. $B_s^0 - \overline{B_s^0}$ mixing in $Z'$ model is also studied by several authors[2,11,36]. But our paper is different from them in the way that we have tried to estimate the mass of $Z'$ boson from $B_q^0 - \overline{B_q^0}$ mass differences.

This paper is organised as follows: In Sec. 2, we discuss the phenomenology of $B_q^0 - \overline{B_q^0}$ mixing (q = d, s) in the Standard Model. In Sec. 3, we discuss about our model and evaluate the mass matrix elements considering contributions from both the Z boson and $Z'$ boson. In Sec. 4, we evaluate the $B_s^0 - \overline{B_s^0}$ and $B_d^0 - \overline{B_d^0}$ mass differences. We summarize our numerical results in Sec. 5.

## 2. $B_q^0 - \overline{B_q^0}$ Mixing in the Standard Model

In the Standard Model, the $B_q^0 - \overline{B_q^0}$ mixing is due to the weak interaction. At the lowest order, this mixing is described by box diagrams involving two W bosons and two up-type quarks (Fig. 1).[4,37] In this case, the long range interactions arising from intermediate virtual states are negligible because the large B mass is off the region of hadronic resonances. In the SM, $M_{12}$ and $\Gamma_{12}$ are computed from the box diagram and read as:[4,38,39]

$$M_{12}^{SM}(B_q) = \frac{G_F^2 M_W^2 M_{B_q} \eta_{B_q}}{12\pi^2} f_{B_q}^2 B_{B_q} S_0(x_t)(V_{tq}^* V_{tb})^2, \qquad (7)$$

$$\Gamma_{12} = \frac{G_F^2 m_b^2 \eta_B' M_{B_q} f_{B_q}^2 B_{B_q}}{8\pi} \times \left[(V_{tq}^* V_{tb})^2 + V_{tq}^* V_{tb} V_{cq}^* V_{cb}\, O\!\left(\frac{m_c^2}{m_b^2}\right) + (V_{cq}^* V_{cb})^2\, O\!\left(\frac{m_c^4}{m_b^4}\right)\right], \qquad (8)$$

where $M_{12}$ and $\Gamma_{12}$ are the off-diagonal elements of the mass and decay matrices, $G_F$ is the Fermi constant, $M_W$ is the W boson mass, $m_i$ is the mass of quark $i$, $x_t = m_t^2/M_W^2$; $M_{B_q}$, $f_{B_q}$ and $B_{B_q}$ are the $B_q^0$ mass, weak decay constant and bag parameter respectively. The Inami – Lim function $S_0(x_t)$ is approximated as $0.784\, x_t^{0.76}$,[40] $V_{ij}$ are the elements of the CKM matrix;[7,8] $\eta_B$ and $\eta_B'$ are QCD corrections.

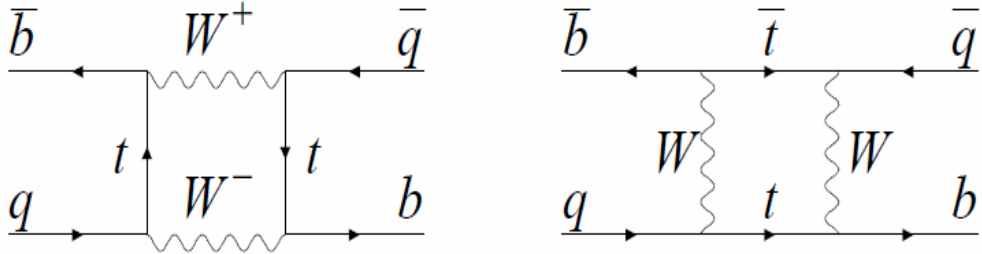

**Fig. 1:** Dominant box diagrams for $B_q^0 - \overline{B_q^0}$ mixing (q = d, s).

The phases of $M_{12}$ and $\Gamma_{12}$ satisfy $\phi_M - \phi_\Gamma = \pi + O\!\left(\frac{m_c^2}{m_b^2}\right),$ \qquad (9)

implying that the mass eigenstates have mass and width differences of opposite signs. The heavy state is expected to have smaller decay width than that of the light state. Hence, $\Delta\Gamma = \Gamma_L - \Gamma_H$ is expected to be positive in the SM.

The quantity $\left|\frac{\Gamma_{12}}{M_{12}}\right| \sim O\left(\frac{m_b^2}{m_t^2}\right)$ is very small. In the absence of CP violation in the mixing, the ratio $\frac{\Delta\Gamma_q}{\Delta M_q}$ is equal to the small quantity $\left|\frac{\Gamma_{12}}{M_{12}}\right|$ which is independent of CKM matrix elements. Hence, it is same for $B_s^0 - \overline{B_s^0}$ and $B_d^0 - \overline{B_d^0}$ systems. From the current experimental knowledge on the mixing parameter $x_q = \Delta M_q / \Gamma_q$,[41] we have

$$x_d = 0.774 \pm 0.008 \quad (B_d^0 - \overline{B_d^0} \text{ system}),$$

$$x_s = 26.2 \pm 0.5 \quad (B_s^0 - \overline{B_s^0} \text{ system}). \quad (10)$$

Furthermore, the Standard Model predicts that $\Delta\Gamma_d / \Gamma_d$ is very small (below 1%), but $\Delta\Gamma_s / \Gamma_s$ is considerably larger (~ 10%).[4] These width differences are caused by the existence of final states to which both the $B_q^0$ and $\overline{B}_q^0$ mesons decay. The $M_{12}^q$ is very sensitive to NP both for $B_d^0$ and $B_s^0$. $\Gamma_{12}^s$ stems from Cabbibo-favoured tree-level decays and possible NP effects are expected to be smaller than the hadronic uncertainties but in the case of $\Gamma_{12}^d$, the contributing decays are Cabbibo-suppressed. New physics in $M_{12}^q$ will not only affect the neutral-meson mixing parameters, but also the time-dependent analyses of decays corresponding to interference between mixing and decay. The $\Delta M_{B_d}$ and $\Delta M_{B_s}$ mass differences in the SM are given in equations (2) and (3).

**3. The Model**

In extended quark sector model[27–29,42], besides the three standard generations of the quarks, there is an $SU(2)_L$ singlet of charge $-1/3$. This model allows for Z-mediated FCNCs. The up quark sector interaction eigenstates are identified with mass eigenstates but down quark sector interaction eigenstates are related to the mass eigenstates by a 4 × 4 unitary matrix, which is denoted by K. The charged-current interactions are described by

$$L_{\text{int}}^W = \frac{g}{\sqrt{2}} \left(W_\mu^- J^{\mu^+} + W_\mu^+ J^{\mu^-}\right), \quad (11)$$

$$J^{\mu^-} = V_{ij} \bar{u}_{iL} \gamma^\mu d_{jL}. \quad (12)$$

The charged-current mixing matrix V is a 3 × 4 submatrix of K :

$$V_{ij} = K_{ij} \quad \text{for } i = 1,......3, \quad j = 1,......,.4. \quad (13)$$

Here, V is parametrized by six real angles and three phases, instead of three angles and one phase in the original CKM matrix.

The neutral-current interactions are described by

$$L_{int}^Z = \frac{g}{\cos\theta_W} Z_\mu \left( J^{\mu 3} - \sin^2\theta_W J_{em}^\mu \right), \tag{14}$$

$$J^{\mu 3} = -\frac{1}{2} U_{pq} \bar{d}_{pL} \gamma^\mu d_{qL} + \frac{1}{2} \delta_{ij} \bar{u}_{iL} \gamma^\mu u_{jL}. \tag{15}$$

In neutral-current mixing, the matrix for the down sector is U = V†V. Since in this case V is not unitary, $U \neq 1$. Its non-diagonal elements do not vanish:

$$U_{pq} = -K_{4p}^* K_{4q} \quad \text{for} \quad p \neq q. \tag{16}$$

Since the various $U_{pq}$ are non-vanishing, they allow for flavour-changing neutral currents that would be a signal for new physics.

Now consider the $B_q^0 - \overline{B_q^0}$ mixing (q = d, s) in the presence of Z-mediated FCNC[27–29,42] at tree level (Fig. 2).[43,44] The Z-mediated FCNC couplings $U_{db}^Z$ and $U_{sb}^Z$, which affect the $B_q^0 - \overline{B_q^0}$ mixing, are constrained from the experimental limit on $B(B \to \ell^+\ell^- X)$.[30–33] The Z-mediated flavour-changing couplings $U_{qb}^Z$ can contribute to $B_q^0 - \overline{B_q^0}$ mixing:[42]

$$M_{12}^Z(B_q) = \frac{\sqrt{2} G_F M_{B_q} \eta_{B_q}}{12} f_{B_q}^2 B_{B_q} (U_{qb}^Z)^2. \tag{17}$$

The same idea can be applied to a Z′-boson i.e., mixing among particles which have different Z′ quantum numbers will induce FCNCs due to Z′ exchange.[26, 45–50] Since the $U_{pq}^Z$ are generated by mixing that breaks weak isospin, they are expected to be at most $O(M_1/M_2)$, where $M_1(M_2)$ is typical light (heavy) fermion mass. On the other hand, the Z′-mediated coupling $U_{pq}^{Z'}$ can be generated via mixing of particles with same weak isospin and, so, suffer no suppression. Even though Z′-mediated interactions are suppressed relative to Z, these are compensated by the factor $U_{pq}^{Z'} / U_{pq}^Z \sim (M_2/M_1)$. Thus, the new contributions from Z′-boson are exactly in the similar manner as in the Z-boson (Fig. 2).[43,44] Therefore, the contribution of Z′-mediated FCNCs to $B_q^0 - \overline{B_q^0}$ mixing[26] is,

$$M_{12}^{Z'}(B_q) = \frac{\sqrt{2} G_F M_{B_q} \eta_{B_q}}{12} \frac{M_Z^2}{M_{Z'}^2} f_{B_q}^2 B_{B_q} (U_{qb}^{Z'})^2. \tag{18}$$

Now considering the contributions from Z- and Z′-mediated FCNC, we can write the mass matrix element for $B_q^0 - \overline{B_q^0}$ mixing as:

$$M_{12}(B_q) = M_{12}^{SM}(B_q) + M_{12}^Z(B_q) + M_{12}^{Z'}(B_q). \tag{19}$$

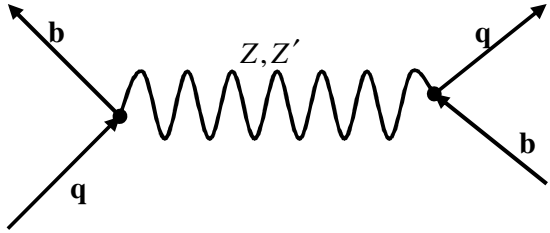

(a)

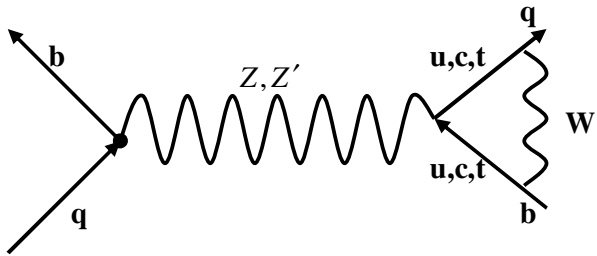

(b)

**Fig. 2:** Feynman diagrams for $B_q^0 - \overline{B_q^0}$ (q = s, d) mixing in the extended quark model, where the blob represents the tree level flavour changing vertex.

# 4. Evaluation of $B_q^0 - \overline{B_q^0}$ mixing mass differences

The $B_q^0 - \overline{B_q^0}$ (q = s, d) mixing mass differences can be evaluated by substituting equations (7), (17), (18) and (19) in equation (1). Thus, considering the contributions from Z- and $Z'$-mediated FCNC, we can write the $B_s^0 - \overline{B_s^0}$ mass difference as:

$$\Delta M_{B_s} = 2 \left[ \frac{G_F^2 M_W^2 M_{B_s} \eta_{B_s}}{12\pi^2} f_{B_s}^2 B_{B_s} S_0(x_t) (V_{ts}^* V_{tb})^2 + \frac{\sqrt{2} G_F M_{B_s} \eta_{B_s}}{12} f_{B_s}^2 B_{B_s} (U_{sb}^Z)^2 \right.$$
$$\left. + \frac{\sqrt{2} G_F M_{B_s} \eta_{B_s}}{12} \frac{M_Z^2}{M_{Z'}^2} f_{B_s}^2 B_{B_s} (U_{sb}^{Z'})^2 \right]$$

……………..(20)

Similarly, the $B_d^0 - \overline{B_d^0}$ mass difference can be written as:

$$\Delta M_{B_d} = 2 \left[ \frac{G_F^2 M_W^2 M_{B_d} \eta_{B_d}}{12\pi^2} f_{B_d}^2 B_{B_d} S_0(x_t) (V_{td}^* V_{tb})^2 + \frac{\sqrt{2} G_F M_{B_d} \eta_{B_d}}{12} f_{B_d}^2 B_{B_d} (U_{db}^Z)^2 \right.$$
$$\left. + \frac{\sqrt{2} G_F M_{B_d} \eta_{B_d}}{12} \frac{M_Z^2}{M_{Z'}^2} f_{B_d}^2 B_{B_d} (U_{db}^{Z'})^2 \right]$$

………(21)

The equations (20) and (21) are used in the next section for our calculations.

# 5. Results and Discussions

We estimate the mass of $Z'$ boson using the experimental values of mass differences i.e. $\Delta M_{B_s} = 17.77 \pm 0.10\,(stat) \pm 0.07\,(syst)\ \text{ps}^{-1}$ [ref. 12] in equation (20) and $\Delta M_{B_d} = 0.507 \pm 0.005\,\text{ps}^{-1}$ [ref. 4] in equation (21). We have taken the recent data from:[4] $G_F = (1.16637 \pm 0.00001) \times 10^{-5}\,\text{GeV}^{-2}$, $M_{B_s} = (5366.0 \pm 0.9)$ MeV, $M_W = (80.399 \pm 0.23)$ GeV, $m_t = 172.0 \pm 0.9 \pm 1.3$ GeV, $M_{B_d} = (5279.5 \pm 0.5)$ MeV, $M_Z = (91.1876 \pm 0.0021)$ GeV. Using the lattice QCD calculations,[51] $f_{B_d}\sqrt{B_{B_d}} = (216 \pm 9 \pm 13)$ MeV, $f_{B_s}\sqrt{B_{B_s}} = (275 \pm 7 \pm 13)$ MeV and assuming $|V_{tb}| = 1$, one finds $|V_{td}| = (8.4 \pm 0.6) \times 10^{-3}$, and $|V_{ts}| = (38.7 \pm 2.1) \times 10^{-3}$. The Inami-Lim function[3] $S_0 = 2.35$, and $\eta_{B_s} = \eta_{B_d} = 0.552$ [ref. 1]. The value of $|U_{sb}^Z| \cong 10^{-3}$ [ref. 52] and $|U_{db}^Z| \cong 10^{-3}$ [ref. 27–29]. From the study of $B_s^0 - \overline{B_s^0}$ mixing in leptophobic $Z'$ model, they[2] obtained $|U_{sb}^{Z'}| \leq 0.036$ for $M_{Z'} = 700$ GeV and $|U_{sb}^{Z'}| \leq 0.051$ for $M_{Z'} = 1$ TeV. We take $|U_{sb}^{Z'}| \approx 0.04$ and $|U_{db}^{Z'}| \approx 7.8 \times 10^{-3}$ for our calculations. With these values, we observe that the value of $\Delta M_{B_s}$ is consistent with the mass of $Z'$ boson in the range 989 GeV – 1665 GeV, which is accessible at the experiments LHC and B-factories in near future.

The contribution of $Z'$-mediated FCNCs to $b \to s \nu \bar{\nu}$ yields the constraint:[26]

$$\left|U_{sb}^{z'}\right|\frac{M_z^2}{M_{z'}^2} \leq 7.1 \times 10^{-3}. \tag{22}$$

We take $\left|U_{sb}^{Z'}\right| \approx 0.04$, and hence our estimation of the mass of $Z'$ boson satisfies the bound of equation (22). This demonstrates the importance of $B_s^0 - \overline{B_s^0}$ mixing in constraining NP in the flavour sector.

Similarly, we take $\left|U_{db}^{Z'}\right| \approx 7.8 \times 10^{-3}$ for our calculations, which is satisfied the constraints obtained for the FCNC coupling $\left|U_{db}^{Z'}\right| \leq 0.61$ for $B \to \pi \nu \bar{\nu}$ decay.[53] We observe that the value of $\Delta M_{B_d}$ is consistent with the mass of $Z'$ boson in the range 1352 GeV – 1824 GeV, which is also accessible at the experiments LHC and B-factories in near future. If one tries with any other values of $Z'$ boson mass, there is a discrepancy in the values of $\Delta M_{B_s}$ and $\Delta M_{B_d}$.

Since the $Z'$ has not yet been discovered, its exact mass is unknown. A broad class of supersymmetric extensions of the SM predict a $Z'$ boson whose mass is naturally in the range 250 GeV $< M_{Z'} <$ 2 TeV.[54] In a study of $B$ meson decays with $Z'$-mediated flavour-changing neutral currents,[47] they study the $Z'$ boson in the mass range of a few hundred GeV to 1 TeV. The current experimental searches of the $Z'$ boson from Drell-Yan cross sections at Tevatron have put lower limits on the mass range 0.6 – 1.0 TeV at 95 % C. L. depending on the specific models.[22] From the electroweak precision data analysis, the improved lower limits on the $Z'$ mass are given in the range 1.1– 1.4 TeV at 95 % C. L..[55] The LHC has the potential of discovering the $Z'$ up to $M_{Z'}$ = 4.5 TeV with 100 fb$^{-1}$ data at center of mass energy $\sqrt{s}$ = 14 TeV.[56] These limits on $Z'$ boson mass favours higher energy ($\geq$ 1 TeV) collisions for direct observation of the signal. It is also possible that the $Z'$ bosons can be much heavy or weak enough to escape beyond the discovery reach expected at the LHC. In this case, only the indirect signatures of $Z'$ exchanges may occur at the high energy colliders.[57] Recently,[58] it has been shown that one can probe a TeV scale $Z'$ boson at the LHC in longitudinal weak gauge boson scattering. More interestingly, our estimation of mass of $Z'$ boson lies in the range of 1352 GeV – 1665 GeV.

In conclusion, the FCNC processes of $B_d^0 - \overline{B_d^0}$ and $B_s^0 - \overline{B_s^0}$ mixing offer interesting probes to search for signals of physics beyond the SM. In this paper, we have tried to estimate the mass of $Z'$ boson from $B_q^0 - \overline{B_q^0}$ mass differences. Our estimation of mass of $Z'$ boson is consistent with the experimental values of $\Delta M_{B_s}$ and $\Delta M_{B_d}$, which is accessible at the experiments LHC and B-factories in near future. Despite of the success of the B-factories and the Tevatron, there is still considerable room for new physics in $B_d^0 - \overline{B_d^0}$ as well as $B_s^0 - \overline{B_s^0}$ mixing. We hope that the current exciting experimental situation will stimulate novel activities in this direction.

**Acknowledgments**

We would like to thank Prof. Amol Dighe, Tata Institute of Fundamental Research, Mumbai, India for helpful discussions and suggestions. We thank the referee for suggesting valuable improvements of our manuscript.